\documentclass[manuscript]{aastex}

\slugcomment{Not to appear in Nonlearned J., 45.}

\shorttitle{Nonlinear Prediction of Solar Cycle 24}
\shortauthors{Kilcik et al.}

\begin{document}


\title{Nonlinear Prediction of Solar Cycle 24}


\author{A. Kilcik\altaffilmark{1}}
\affil{Department of Physics, Akdeniz University, Faculty of Arts and Sciences, 07058 Antalya, Turkey.}
\email{alikilcik@akdeniz.edu.tr}

\author{C.N.K. Anderson \altaffilmark{2}}
\affil{Scripps Institution of Oceanography, University of California, San Diego 92103 USA.}
%\email{senka@UCSD.Edu}

\author{J.P. Rozelot \altaffilmark{3}}
\affil{Universit\'e de Nice Sophia-Antipolis CNRS-OCA, Fizeau Dpt, Av. Copernic, 06130 Grasse, France.}
%\email{Jean-Pierre.Rozelot@obs-azur.fr}

\author{H. Ye \altaffilmark{2}}
\affil{Scripps Institution of Oceanography, University of California, San Diego 92103 USA.}
%\email{hye@ucsd.edu}

\author{G. Sugihara \altaffilmark{2}}
\affil{Scripps Institution of Oceanography, University of California, San Diego 92103 USA.}
%\email{gsugihara@UCSD.Edu}

\and

\author{A. Ozguc \altaffilmark{4}}
\affil{Kandilli Observatory and E. R. I., Bogazici University, Cengelkoy, Istanbul, Turkey.}
%\email{ozguc@boun.edu.tr}

%\altaffiltext{1}{Visiting Astronomer, Cerro Tololo Inter-American Observatory.
%CTIO is operated by AURA, Inc.\ under contract to the National Science
%\altaffiltext{2}{Society of Fellows, Harvard University.}
%\altaffiltext{3}{present address: Center for Astrophysics,
%    60 Garden Street, Cambridge, MA 02138}
%\altaffiltext{4}{Visiting Programmer, Space Telescope Science Institute}
%\altaffiltext{5}{Patron, Alonso's Bar and Grill}


\begin{abstract}
Sunspot activity is highly variable and challenging to forecast. Yet forecasts are important, since peak activity has profound effects on major geophysical phenomena including space weather (satellite drag, telecommunications outages) and has even been correlated speculatively with changes in global weather patterns. This paper investigates trends in sunspot activity, using new techniques for decadal-scale prediction of the present solar cycle (cycle 24). First, Hurst exponent $H$ analysis is used to investigate the autocorrelation structure of the putative dynamics; then the Sugihara-May algorithm is used to predict the ascension time and the maximum intensity of the current sunspot cycle. Here we report $H$ = 0.86 for the complete sunspot number dataset (1700-2007) and $H$ = 0.88 for the reliable sunspot data set (1848-2007). Using the Sugihara-May algorithm analysis, we forecast that cycle 24 will reach its maximum in December 2012 at approximately 87 sunspots units. 
\end{abstract}





\section{Introduction}

Solar radiation is far from constant. These changes can be seen in many solar activity indicators, such as sunspot number, sunspot area, total solar irradiance, solar flares, the recurrence index of geomagnetic disturbances \citep{kan08}, dynamo models \citep{hir01}, and solar cycle length \citep{kan08}. However, sunspot number is the most common solar activity indicator, having been recorded since 1700, and used as an indicator for solar cycle prediction since 1913 \citep{kim13, cur73, dem81, kan07b}.   
\par Any change in solar activity presents challenges in solar physics (understanding solar cycle mechanisms, prediction of solar events such as flares, CME, etc) or in the field of space weather (satellite drag, telecommunication outages, etc). Therefore solar cycle prediction is of vital importance now, and will be even more important in the future. Despite general knowledge of solar
cycles, reliable forecasting of sunspot number remains problematic. Several statistical methods have already been applied, but due to the nonlinear nature of the time series, linear approaches are expected to fail. Other techniques have been employed (e.g. precursors, \citep{kan07a}) that offer theoretical advantages. Because of the wide variety of techniques used, the present solar cycle (Cycle 24) has been predicted to have a maximum sunspot number as low as 50 or as high as 180 \citep{obr08}. In this study, we predict the length and intensity of cycle 24 using two statistical/dynamical methods: the Hurst exponent and the Sugihara-May algorithm.

\section{Data and Methods}

%% In a manner similar to \objectname authors can provide links to dataset
%% hosted at participating data centers via the \dataset{} command.  The
%% second curly bracket argument is printed in the text while the first
%% parentheses argument serves as the valid data set identifier.  Large
%% lists of data set are best provided in a table (see Table 3 for an example).
%% Valid data set identifiers should be obtained from the data center that
%% is currently hosting the data.
%%
%% Note that AASTeX interprets everything between the curly braces in the 
%% macro as regular text, so any special characters, e.g. "#" or "_," must be 
%% preceded by a backslash. Otherwise, you will get a LaTeX error when you 
%% compile your manuscript.  Special characters do not 
%% need to be escaped in the optional, square-bracket argument.
The monthly ISSN (International SunSpot Number) data were used from 1847 to 2007. Though data is available back to 1700,  according to \citet {kan08} ``the quality of the
data is considered as poor during 1700 - 1748, questionable during 1749 - 1817, good during
1818 - 1847, and reliable since 1848.'' The maximum of intensity of each cycle and the time from the
beginning of the cycle to its maximum (hereafter ``ascension time'') was derived for cycles  1 (February 1755) to 23 (March 2000). All the datasets used in this study were taken from National Geophysical Data Center (NGDC)\footnote{http://www.ngdc.noaa.gov/stp/SOLAR/ftpsunspotnumber.html}. 
Because standard methods are expected to fail to make accurate predictions, we apply two unusual techniques to this data to make our predictions.
\par (1) The Hurst Exponent or Rescaled Range Analysis. This method was proposed by \citet{hur51} for an experimental study of long term information storage in time series data. This technique has had wide application in many research fields, including finance \citep{gre04,qia04}, astronomy \citep{kom95,roz95,roz08,ruz94}, climate \citep{ran04}, and others. Hurst analysis is a simple and robust way to analyze randomness in a dataset. The parameter $H$ measures the persistence of structures in the time series, indicating whether the data represent a pure random walk or have underlying trends. Another way to state this is that a random process with an underlying trend has some degree of autocorrelation. When the autocorrelation has a very long (or mathematically infinite), decay the process is referred to as a long memory process. The value of $H$ varies from 0 (indicating anti-persistent brown noise) to 0.5 (random white noise), 1.0 (indicating a strong, smooth trend). Hurst found that the rescaled range series $(R / S)$ over a time window of width $t$ is described as a power law:
\begin{equation}
(R / S)_{t}=c^{*}t^{H},
\end{equation}
where $c^{*}$ is a constant and $H$ is the Hurst exponent. To estimate the value of the Hurst exponent, $R/S$ is plotted versus $t$ on log-log axes. The slope of the linear regression gives the value of the Hurst exponent. More details about $R/S$ analysis, see \citet{qia04}. 
\par (2) The Sugihara-May Algorithm. 
This algorithm compares a library of (known) past patterns to patterns seen later in the real time series. It does this by reconstructing an attractor from the library, locating the ``present'' point on the attractor, and tracing that point forward along the attractor's trajectory. The dimension of the attractor is determined by the complexity of the process that generates it, and the number of points involved in tracing forward determines the nonlinearity of the system. A similar chaos technique was firstly applied to the historical sunspot data by \citet{kur90} to determine the nonlinearity of the data set and to predict the following solar cycle. For a chaotic time series, the accuracy of nonlinear forecast falls off as prediction time increases \citep{sug90}. This is a two step procedure. First, simplex projection identifies the best embedding dimension, which is then used in a S-map procedure to check the nonlinearity \citep{sug90,sug94}. A time series $X$ of length $N$, is embedded in $D$-dimensional Euclidean space to create a ``landscape'' $x$ of $N - D - 1$ vectors, where {\it {$x_{t}$: ($X_{t-D+1}, X_{t-D+2},... , X_{t}$)}}. The first $n$ of these {\it{$x_{t}$}} vectors are associated with output values {\it{$X_{t+Tp}$}}. Then forecasts of the $(N - n)$ remaining input vectors, i.e. predictions, are made by  

\begin{equation}
\hat{y}_{t+Tp}=\frac{\sum^{D+1}_{k=1}X_{k+Tp}exp(-d_{k})}{\sum^{D+1}_{k=1}exp(-d_{k})}
\end{equation} 
\begin{equation}
d_{k}=\left\|y_{t}-x_{k}\right\|
\end{equation}
where the summation is taken over the $D+1$ closest neighbors in $D$-dimensional Euclidean space. Ideally, these closest neighbors form the vertices of the smallest simplex containing the predictee. In this context, this predictive approach is a local approximation similar to kernel density estimation. To find an appropriate value of D, various values of D are applied to find the value that minimizes prediction error (i.e., an embedding that minimizes the singularities or indeterminate crossing points of trajectories in the putative attractor). The optimal $D$ is then used in an S-map to build weighted (local) linear predictions. The rate of decay of weight given to each point is set by $\theta$ parameter, and it describes the degree of caoticity of a data set. The forecast improvement with  local weighting indicates that the underlying dynamics are nonlinear. Hence $\theta$ = 0 performs best for linear time series (i.e., $\theta$ = 0 is the global linear solution), and when $\theta$ $>$0 performs best, the time series is nonlinear \citep{miy00}. 
\par We investigate the significance of the historical relationship between maximum sunspot number and ascension time with Pearson's test.
%% In this section, we use  the \subsection command to set off
%% a subsection.  \footnote is used to insert a footnote to the text.


%% This section also includes several of the displayed math environments
%% mentioned in the Author Guide.

\section{Analysis and Results}
Hurst exponent analysis was applied to the monthly ISSN data separated in two time intervals: 1700 to 2007 and 1848 to 2007. $H$= 0.86$\pm$0.008, and $H$=0.88$\pm$0.009 for these datasets respectively (Fig. 1). The bulge between 6 and 8 in panels (a) and (b) of Figure 1 is due to the main contribution of the 11-year cycle \citep{ruz94}.
\par To investigate the complexity, nonlinearity and predictability of the data, the Sugihara-May algorithm was applied as described above. From this analysis, it was discovered that the
data are high-dimensional, i.e. governed by many orthogonal processes such as solar rotation and inner
magnetic activity (Fig. 2).
\par In the second step of this procedure, the S-Map analysis (Fig. 3), a weak nonlinear signature was detected in 8 dimensions, indicating weak chaotic behavior. 
\par We also tested the prediction capability of our method by predicting the second half of the given data set (1927-2008) using only data from 1848-1926. The model produced very skillful predictions, showing a correlation of 0.94 with the observed values (Fig. 4). We therefore attempted to perform long-range forecasts by sequentially increasing Tp (the time length into the future that one tries to predict from the last data point). These results are mostly an echo of the previous cycle. This technique predicts that the current solar cycle will reach a maximum on December 2012, peaking at a sunspot number of 87.4 (Fig. 5).   
\par A similar method to our prediction analysis was used by \citet{kur90} for the solar cycle 22. They found that the maximum sunspot number would be about 150. Because the observed maximum of that cycle was 158, a rather good prediction estimate, it instills more confidence in the technique.
\par For historical cycles, we further investigated the link between the maximum sunspot number and the ascension time. We found that the two variables have a strong negative correlation (r=0.82, df=21, and p$<$.001; Fig. 6). Our prediction above (87.4 sunspots in 5.1 years from December 2007) is plausible given the historical relationship between the two values, as cycles with an ascending phase between 4.5 and 5.5 years have a maximum in the 64.2 to 131.6 sunspot range. 

\section{Discussion}
Our results show that the Hurst exponents of the ISSN data for the periods of time 1700-2007 and 1848-2007 are 0.86$\pm$0.008 and 0.88$\pm$0.009 respectively. This is in agreement with \citet{man69} who first applied this type of analysis to the monthly sunspot number, and found $H$ = 0.86. Similarly  \citet{ruz94} analyzed the $^{14}C$ radiocarbon data as a proxy of solar activity, and found $H$= 0.84. In analyzing daily Doppler solar differential rotation coefficients A and B measured at Mount
Wilson, USA, \citet{kom95} found $H$ to be 0.83 and 0.86 respectively. The slightly higher exponent obtained for the recent dataset indicates that more reliable sunspot data exhibits a slightly stronger autocorrelation (or tendency to trend), and thereby has more statistical predictability than the full data set (1700-2007). Thus it is likely that errors or uncertainty within the unreliable data affected its inherent predictability.
\par Our prediction of ascension time compare with values already available in the literature, which range from December 2009 to 2014 \citep{mar06, tsi97}. \citet{obr08} reported that the second half of 2010 or the first quarter of 2011 would be the most reliable extimates for the maximum of cycle 24. \citet {kan08} suggested October 2011 or August of 2012, while \citet {mar06} found that cycle maximum will occur as early as December 2009. 
\par Our prediction of Cycle 24's intensity also bears comparison with other forecasts. As early as 1983, \citet {chi83} claimed that cycle 23 and 24 would be low and cycle 24 would be lower than cycle 23. \citet{duh03} predicted a maximum of 87.5$\pm$23.5, and \citet{wan02} estimated a peak between  83.2 and 119.4. \citet {mar04} reported that cycle 24 has to be low by analyzing the number of flares. \citet {jav07} obtained a value of 74 from the sunspot group data and \citet {hir07} obtained a value of 116 by using a harmonic oscillator solar cycle model. \citet {kan07a} predicted that sunspot number would be 129.7$\pm$16.3 for the present Cycle, but later (2008) revised this estimate to either 140$\pm$20 in October 2011 or 90$\pm$10 in August 2012. It must be emphasized that the observational result for the start of cycle 24 (April 2008), is in rather good agreement with Kane's prediction. Thus, we can
think that Kane's estimates for the time at which cycle 24 will reach its maximum could be reliable. Using neuronal prediction, \citet {mar06} predicted 145 sunspots at the peak in December 2009.
\par The precursor models, which use data from the declining phase of cycle $N-1$ to predict height and timing of cycle $N$, were used by \citet{obr08}. They reported that ``the precursor models based on the polar field or $H_{\alpha}$ data often yield lower values of the current cycle'', which vary from 70$\pm$10 and 120$\pm$40. From the dynamo model \citet{cho07} have reported that cycle 24 would be 35$\%$ lower than cycle 23. \citep{dik06} disagreed--using a flux transport dynamo model they argued that cycle 24 will be 30 - 50$\%$ higher than cycle 23. \citet{hat06} give a value of 160 based on geomagnetic activity at the minimum. From the index of the global magnetic field, \citet{obr08} forecast that cycle 24 would be of ``medium high, the same or somewhat higher than cycle 23''. Finally, \citet {kit08} using a nonlinear dynamo model as described by \citet {kle82} which
takes into account dynamics of the turbulent magnetic helicity, predict that the next sunspot cycle will be significantly weaker (by $\sim 30\%$) than the previous cycle, continuing the trend of low solar activity.

\par By means of the two methods utilized here, the Rescaled Range Analysis and the Sugihara-May algorithm, we were able to deduce from the significant trends of the ISSN data, both the cycle duration and its maximum intensity. Unlike previous analysis our forecasting results are based on models that were tested out of sample to have a high degree of forecast skill. Our conclusions therefore are (i) the reliable monthly mean sunspot data during 1848-2007 \citep{kan08} yield a slightly higher Hurst exponent than do all the historical observational data; (ii) $H$, being greater than 0.5, shows that the sunspots series are highly persistent (exhibit momentum or trending); (iii) concerning cycle 24, the maximum intensity of 87.4 will be reached in December 2012; (iv) according to this forecast, the current solar cycle will have a magnitude far lower than any other since 1890-1910.

\acknowledgments
All data sets used in this study are taken from NGDC web page. One of the authors (A. K.) is very thankful to the SWYA School organizing committee for providing financial support and to the lecturers for their valuable comments.  This work which is a small part of Ph D thesis of A.K., was supported by the Scientific and Technical Council of Turkey by the project of 107T878. 


%% in parentheses, after the keyword, but they will not be verified.



\clearpage

%% the file, uncomment the \plotone, \plottwo, and
%\includegraphics commands
%%
%% If you need a layout that cannot be achieved with \plotone or
%% \plottwo, you can invoke the graphicx package directly with the
%% \includegraphics command or use \plotfiddle. For more information,
%% please see the tutorial on "Using Electronic Art with AASTeX" in the
%% documentation section at the AASTeX Web site,
%% http://www.journals.uchicago.edu/AAS/AASTeX.
%%
%% The examples below also include sample markup for submission of
%% supplemental electronic materials. As always, be sure to check
%% the instructions to authors for the journal you are submitting to
%% for specific submissions guidelines as they vary from

%% This example uses \plotone to include an EPS file scaled to
%% 80% of its natural size with \epsscale. Its caption
%% has been written to indicate that additional figure parts will be
%% available in the electronic journal.

%\begin{figure}
%\epsscale{.80}
%\plotone{fig1.eps}
%\caption{Monthly mean sunspot numbers from the beginning of 1848 to the end of 2007.}
%\end{figure}

%% Here we use \plottwo to present two versions of the same figure,
%% one in black and white for print the other in RGB color
%% for online presentation. Note that the caption indicates
%% that a color version of the figure will be available online.
%%
\begin{figure}
\epsscale{.80}
\plotone{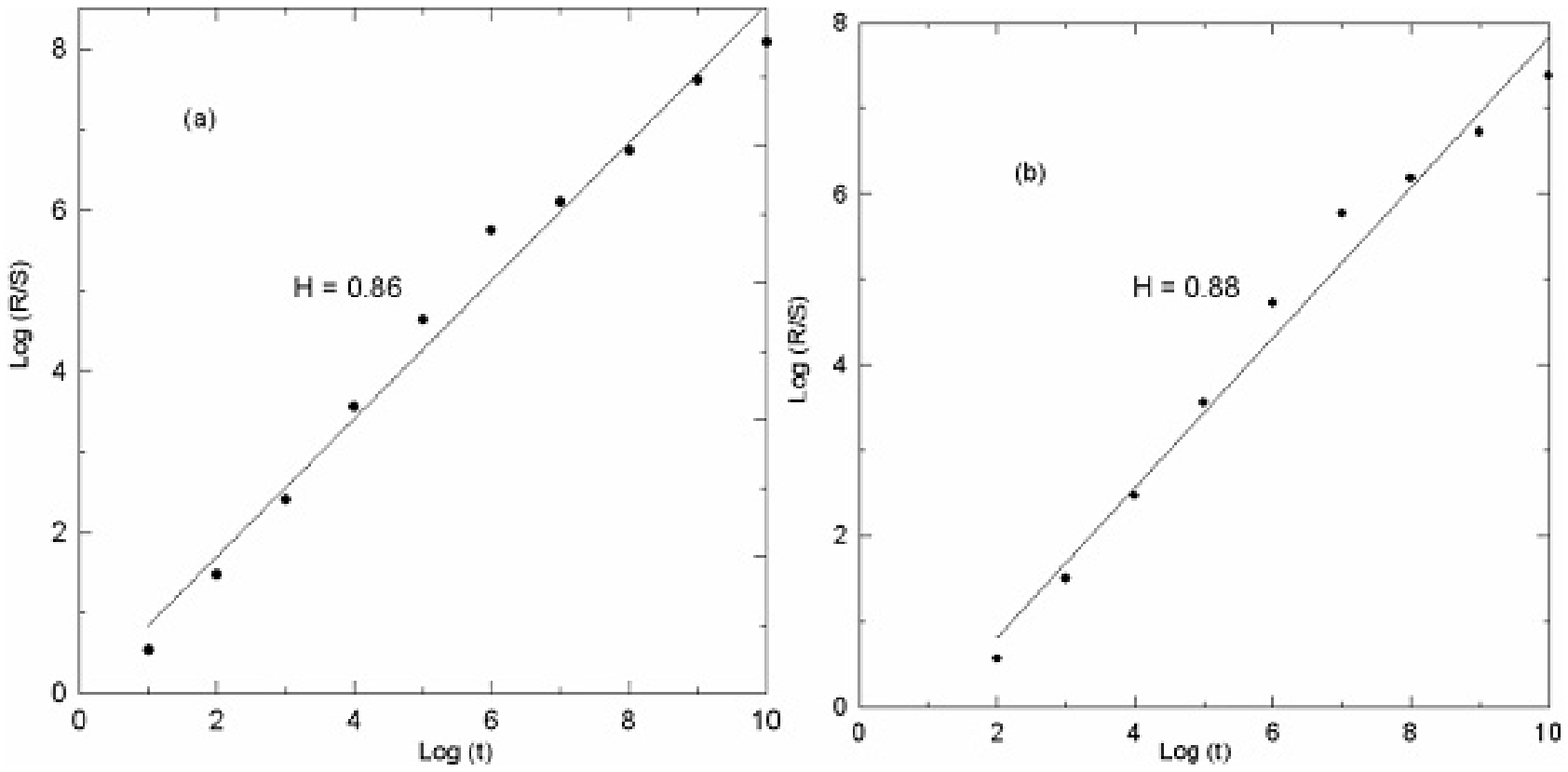}
\caption{Hurst exponent analysis results deduced from monthly ISSN data sets; a) from 1700 to 2007, b) from 1848 to 2007. The slight departure from linearity is a signature of the cyclicity.}
\end{figure}

%\clearpage

\begin{figure}
\epsscale{.80}
\plotone{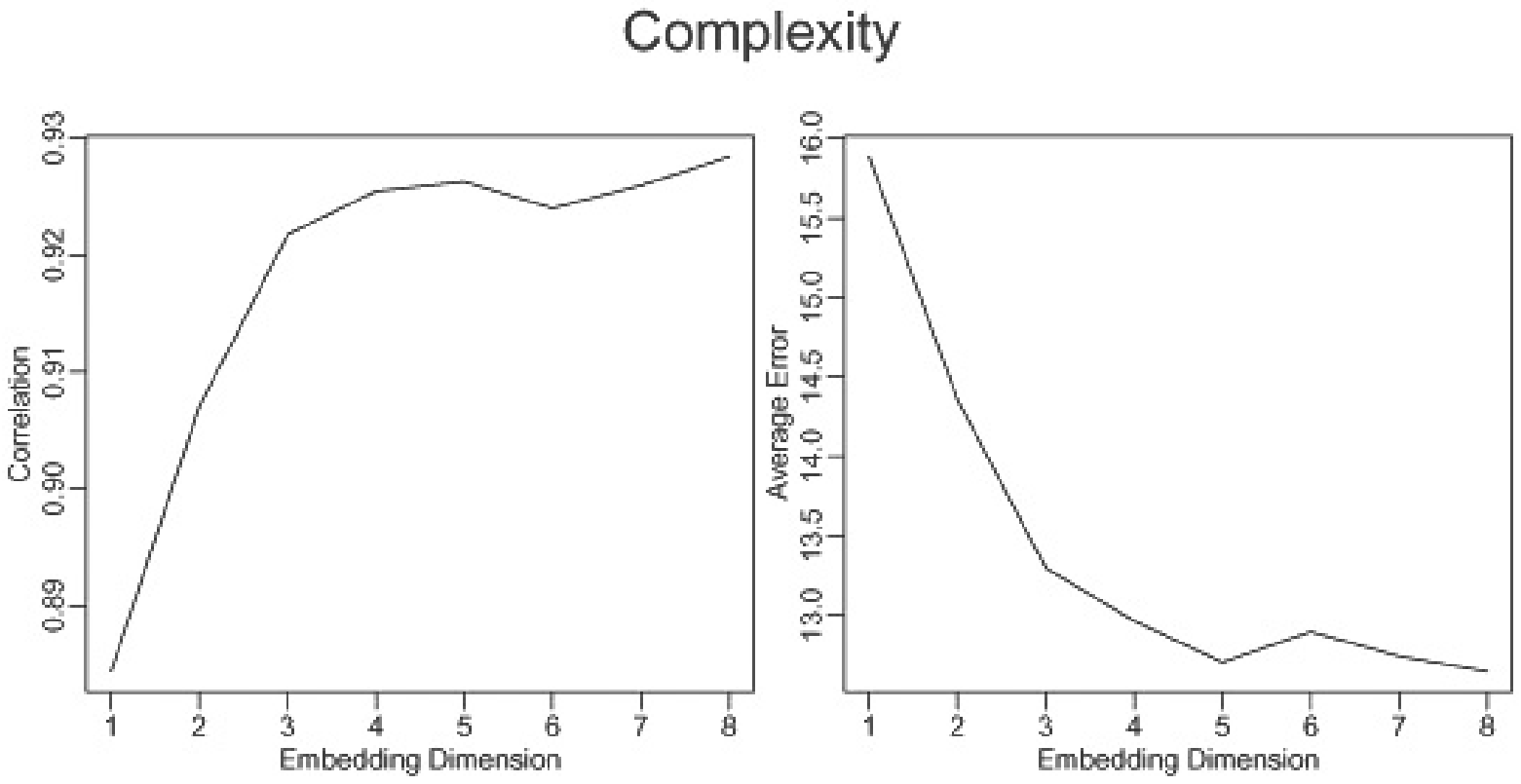}
\caption{Complexity of the ISSN data. These figures show that the best embedding dimension is moderate (D=5-8).}
\end{figure}

\begin{figure}
\epsscale{.80}
\plotone{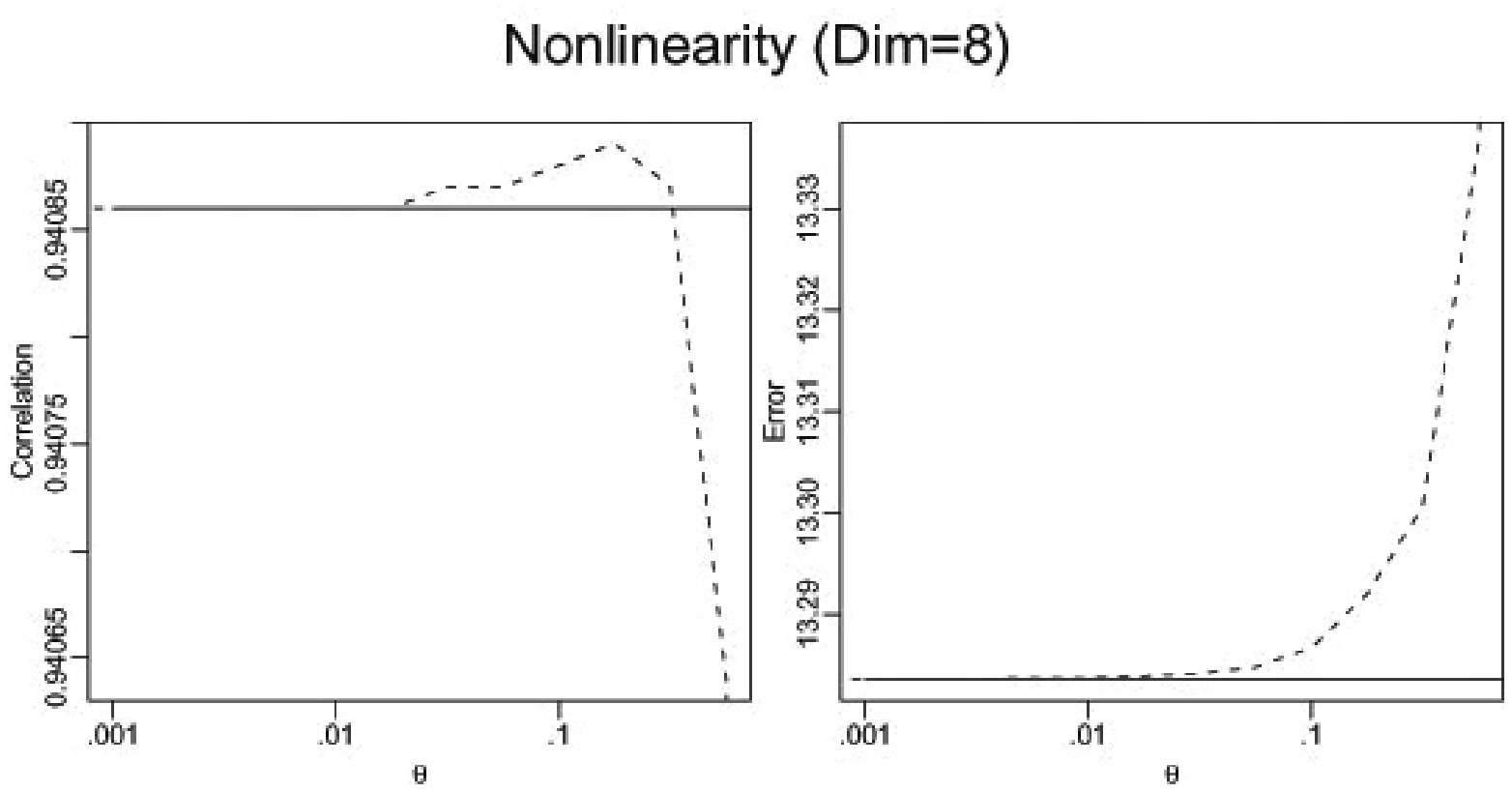}
\caption{For the best embedding dimension of (D=8) as deduced from Fig. 2, the data set is most predictable when using a nonlinearity tuning $\theta > 0$, of about 0.18. For linear data sets this value is equal to zero. Qualitatively similar results are obtained with D=5.}
\end{figure}

\begin{figure}
\epsscale{.80}
\plotone{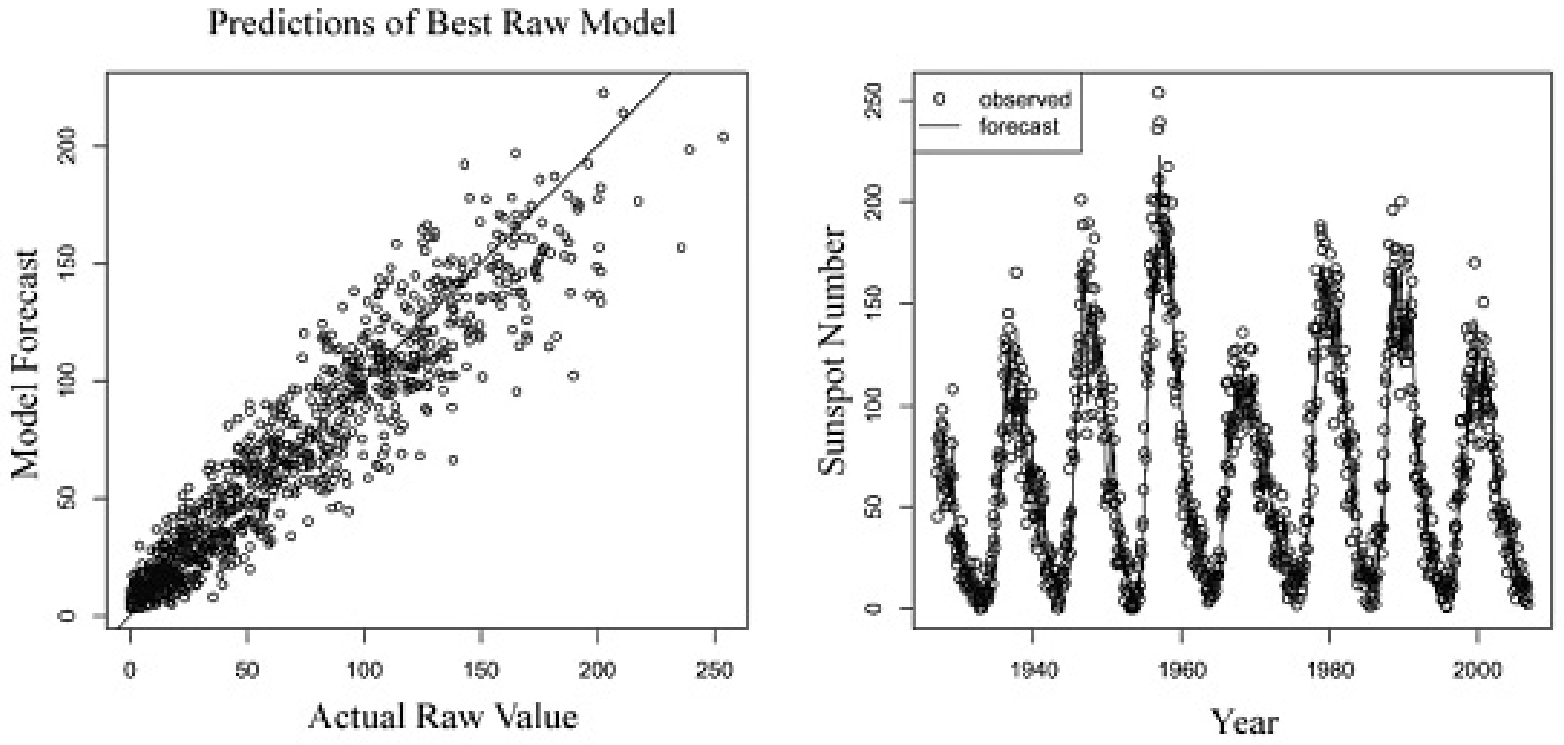}
\caption{(a) The correlation / error between observed and real data. (b) Comparison of observed and predicted ISSN values for the used data period (1848-2007).}
\end{figure}

%% This figure uses \includegraphics to scale and rotate the still frame
\begin{figure}
\epsscale{.80}
\plotone{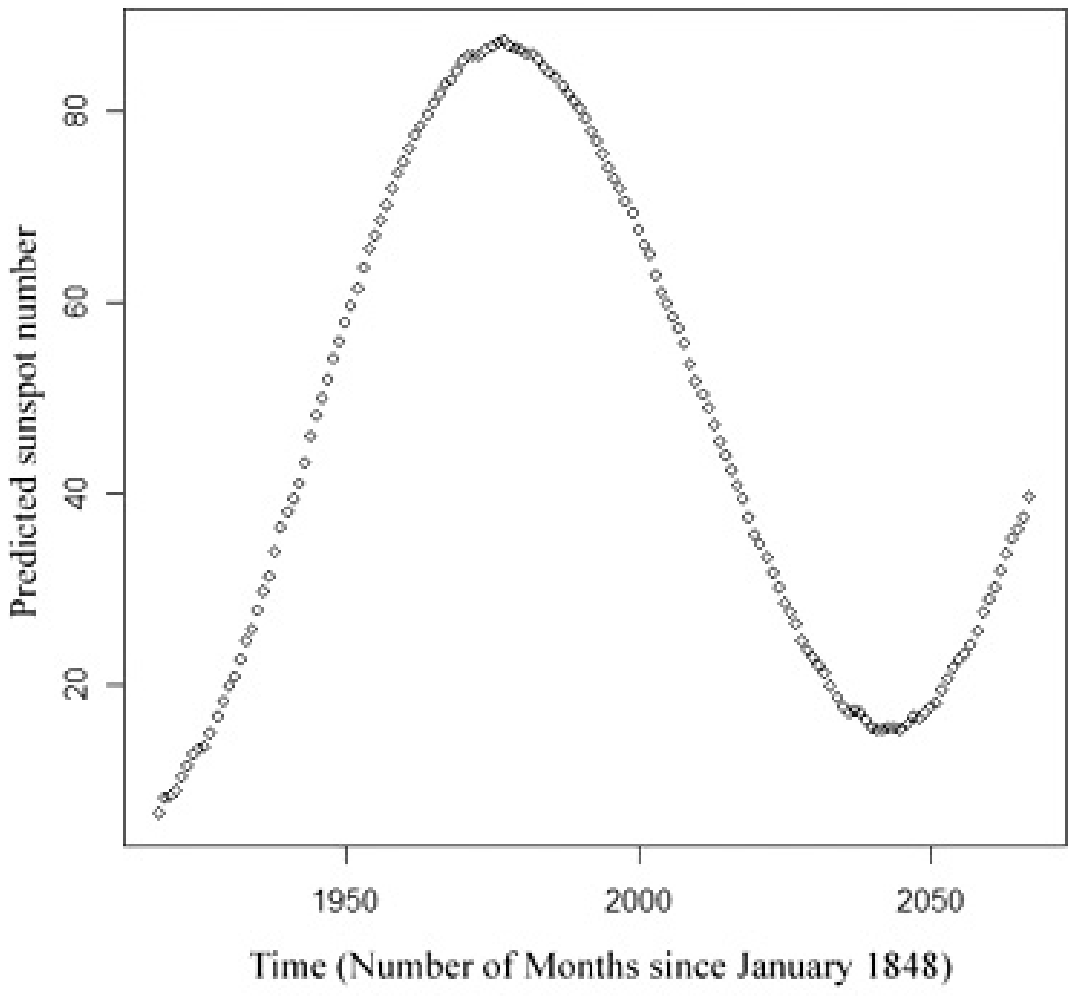}
\caption{Monthly ISSN data predictions by using Sugihara and May Algorithms. The first value is taken as 1918 which describes also January 2008.}
\end{figure}

\begin{figure}
\epsscale{.80}
\plotone{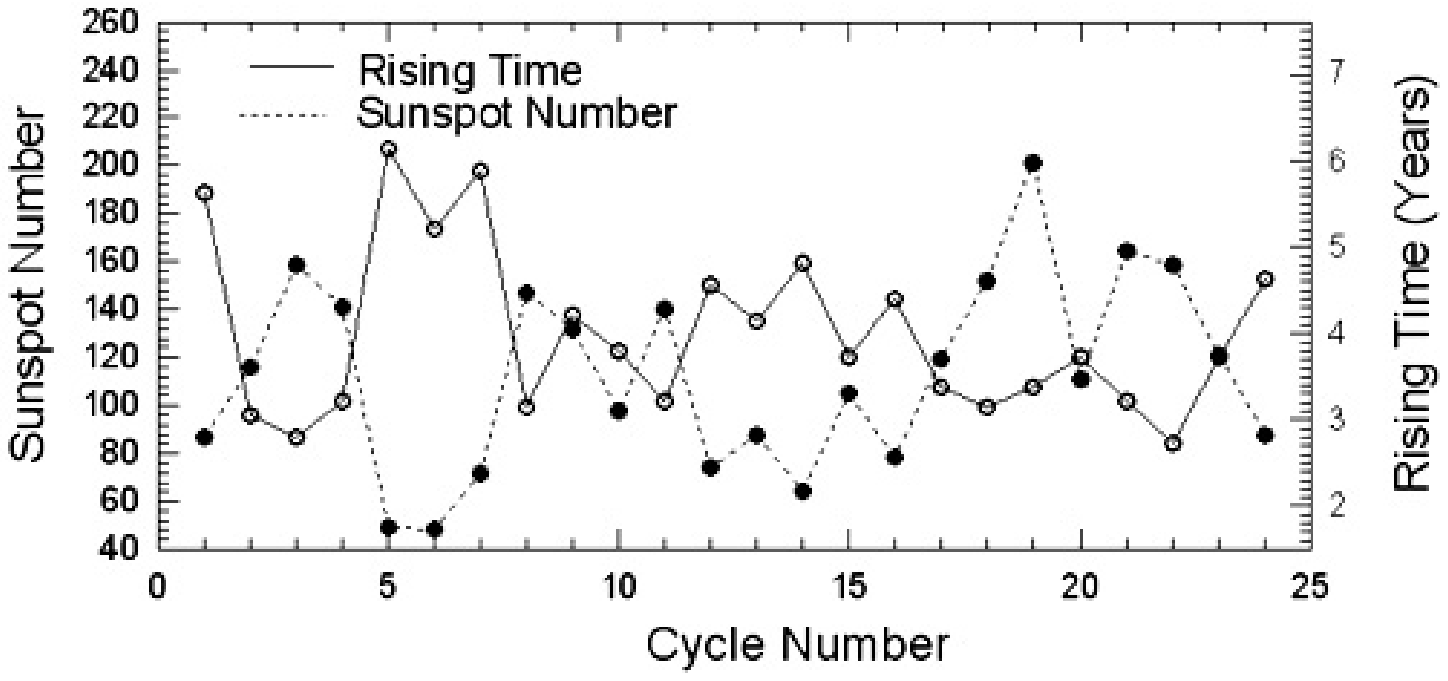}
\caption{Comparisons of sunspot numbers and ascension time of historical cycles (r=0.82, df=21, and p$<$.001). The two last values describe our prediction.}
\end{figure}

\end{document}